\documentclass[aps,prc,superscriptaddress,nofootinbib,showpacs,floatfix,singlecolumn,longbibliography]{revtex4}
\usepackage{url}
\usepackage{cancel}
\usepackage[colorlinks,linkcolor=blue,citecolor=blue,filecolor=black,urlcolor=blue]{hyperref}
\usepackage{epsfig,graphics}
\usepackage{graphicx}
\usepackage{bm}
\usepackage[usenames]{color}
\usepackage{amssymb}
\usepackage{amsmath}
\usepackage{multirow}
\usepackage{float}
\usepackage{harpoon}
\usepackage{MnSymbol}
\usepackage{appendix}
\usepackage{color}
\usepackage{hyperref}
\usepackage{cleveref}
\usepackage{lineno}
\usepackage{lipsum}
\usepackage{ulem}
\newcommand{\Nch} {N_{\mathrm{ch}}}
\newcommand{\sqn}{\sqrt{s_{\mathrm{NN}}}}

\newcommand{\lr}[1]{\left\langle #1\right\rangle}
\newcommand{\llrr}[1]{\llangle #1\rrangle}
\newcommand{\pT}{\mathrm{p_{\mathrm{T}}}}
\usepackage{CJK}
\begin{document}
\title{Higher-order transverse momentum fluctuations in heavy-ion collisions}
\newcommand{\sbu}{Department of Chemistry, Stony Brook University, Stony Brook, NY 11794, USA}
\newcommand{\bnl}{Physics Department, Brookhaven National Laboratory, Upton, NY 11976, USA}
\author{Somadutta Bhatta}\affiliation{\sbu}
\author{Chunjian Zhang}\affiliation{\sbu}
\author{Jiangyong Jia}\email[Correspond to\ ]{jiangyong.jia@stonybrook.edu}\affiliation{\sbu}\affiliation{\bnl}
\date{\today}
\begin{abstract}
In relativistic heavy-ion collisions, the event-by-event mean transverse momentum fluctuations are sensitive to overlap area and energy density fluctuations in initial state. We present a framework to calculate $p_{\mathrm{T}}$ fluctuations up to $4^{\mathrm{th}}$-order using standard and subevent methods, which is validated using the HIJING model. We observe a power-law dependence for cumulants of all orders as a function of charged particle multiplicity $N_{\mathrm{ch}}$, consistent with a simple independent source picture. The fluctuation in $pp$ collisions is observed to be larger than for $p+$Pb, Pb+Pb and Xe+Xe collisions at the same $N_{\mathrm{ch}}$ due to bias in number of contributing sources. The short-range correlations are greatly suppressed in subevent method in comparison to calculations based on the standard method. This study provides a baseline for transverse momentum fluctuations without the presence of final state effects.
\end{abstract}
\maketitle
\section{Introduction}\label{sec:intro}
Heavy-ion collisions at ultrarelativistic energies creates a hot and dense QCD matter whose space-time evolution is well described by relativistic viscous hydrodynamics~\cite{Romatschke:2017ejr,Busza:2018rrf}. Fluctuations of the position of participating nucleons lead to large event-by-event fluctuations in the initial state of the collision. As the system expands, initial-state fluctuations survive various stages of system evolution and influence the final-state observables~\cite{Alver:2010gr,Teaney:2010vd,Schenke:2020uqq}. Within hydrodynamic model framework, the average transverse momentum $\pT$ in each event depends on the initial energy density and is inversely related to the size of overlap region area~\cite{Shuryak:1997yj,Gardim:2019xjs, Giacalone:2020lbm,Jia:2021qyu}. The variance of the $\pT$ and the initial overlap size $R$ is expected to have a simple relation $\sigma(\pT)/\lr{\pT} = \sqrt{\big \langle(\pT-\lr{\pT})^{2}\big \rangle}/\lr{\pT} \propto \sigma(R)/\lr{R}$~\cite{Bozek:2012fw}. The $\pT$ fluctuations can be quantified by its cumulants such as mean, variance, skewness and kurtosis. But so far, only the mean and variance have been measured experimentally~\cite{STAR:2005vxr,ALICE:2013rdo, ALICE:2014gvd, STAR:2017sal, STAR:2019dow,  ATLAS:2019pvn}. Recently Ref.~\cite{Giacalone:2020lbm} predicted that the skewness of $\pT$ fluctuations should be significantly larger than expectation from the independent source scenario, and they can be used to probe nuclear deformation effects~\cite{Jia:2021qyu,Jia:2021tzt}. But no experimental measurement has been carried out for skewness and kurtosis.

The cumulants of $\pT$ fluctuations are intensive quantities. Within the independent source picture such as that implemented in the HIJING model, each collision comprises of superposition of independent $pp$-like collisions and interaction between the sources are ignored~\cite{Wang:1991hta}. Under these conditions, the $n^{\mathrm{th}}$-order cumulant is expected to scale as $\propto 1/\mathrm{N_{s}}^{(n-1)}$, where $\mathrm{N_{s}}$ is the number of sources often taken to be $N_{\mathrm{part}}$ (number of participating nucleons) or $\Nch$ (charged particle multiplicity)~\cite{Zhou:2018fxx,Giacalone:2020lbm,Cody:2021cja}. However, the dynamics of evolution and final-state interaction can lead to a deviation from this power-law behavior. Indeed, experimental measurements of $\pT$ variance in Au+Au collisions at $\sqn$ = 200 GeV and Pb+Pb collisions at $\sqn$ = 2.76 TeV have reported the power to be $\sim0.81$ instead of the expected value of 1. This clear deviation from the baseline of independent source picture in the experimental data indicates the presence of long-range collective correlations and significant final-state effects~\cite{ALICE:2014gvd, ATLAS:2017rtr,STAR:2019dow}.

The direct calculation of higher order cumulants in heavy-ion collisions are computationally expensive, we provide a framework based on the standard and subevent methods to ease this calculation\cite{Bilandzic:2010jr,Jia:2017hbm}. We use the HIJING model to extract the baseline in which $\pT$ fluctuations arise mostly from superposition of contributions from independent sources. The contribution from each source contains both long-range correlations associated with strings and short-range correlations from jets and resonance decays~\cite{Wang:1991hta}. The short-range correlations can be suppressed using rapidity-separated subevent methods, which are widely used in in previous flow analyses~\cite{ATLAS:2017rtr, Jia:2017hbm, Huo:2017nms,ATLAS:2018ngv,CMS:2019lin,ATLAS:2019peb,ATLAS:2020sgl}. To test the power-law scaling and its system size dependence, we repeated the study using Pb+Pb, Xe+Xe $p$+Pb and $pp$ collisions at similar center of mass energy around 5 TeV.

The paper is organised as follows. Section~\ref{sec:method} provides the formulae for calculating $\pT$ cumulants in both standard and subevent methods and describe the setup for HIJING model. The main results are presented in Section~\ref{sec:result}. Section~\ref{sec:summary} gives a summary.

\section{Methodology and Model Setup}\label{sec:method}
We first provide the framework for mean-$\pT$ cumulants following the approach prescribed in Refs.~\cite{DiFrancesco:2016srj,Jia:2017hbm}. The $n$-particle $\pT$ correlator in one event is defined as:
\begin{align}\label{eq:ck}
c_{n}=\frac{\sum_{i_1\neq ...\neq i_n } w_{i_1}...w_{i_n}(p_{\rm{T},i_1}-\llrr{\pT})...(p_{\rm{T},i_n}-\llrr{\pT})} { \sum_{i_1\neq ...\neq i_n } w_{i_1}...w_{i_n}}
\end{align}
where $w_i$ is the weight for particle $i$. Following~\cite{Jia:2017hbm} and denoting $p\equiv \pT$, this relation can be expanded algebraically into a simple polynomial function of the following quantities:
\begin{align}
& p_{m k}=\sum_{i}w_{i}^{k}p^{m}_{i} / \sum_{i}w_{i}^{k},   \tau_{k}=\frac{\sum_{i} w_{i}^{k+1}}{\left(\sum_{i} w_{i}\right)^{k+1}} \notag \\
& \bar{p}_{1 k} \equiv p_{1 k}-\llrr{\pT} \notag \\
& \bar{p}_{2 k} \equiv p_{2 k}-2p_{1 k}\llrr{\pT}+\llrr{\pT}^{2} \notag \\
& \bar{p}_{3 k} \equiv p_{3 k}-3p_{2 k}\llrr{\pT}+3p_{1 k}\llrr{\pT}^{2}-\llrr{\pT}^{3} \notag \\
& \bar{p}_{4 k} \equiv p_{4 k}-4p_{3 k}\llrr{\pT}+6p_{2 k}\llrr{\pT}^{2}-4p_{1 k}\llrr{\pT}^{3}+\llrr{\pT}^{4} \label{eq:pmk}
\end{align}
Note there $\llrr{\pT}=\lr{p_{11}}$ is the mean-$\pT$ averaged over the event ensemble.

Using these auxiliary variables, the correlator in Eq.~\ref{eq:ck} in the standard method can be expressed as,
\begin{gather}
c_{2}=\frac{\bar{p}_{11}^2-\tau_{1}\bar{p}_{22}}{1-\tau_{1}} ,\quad
c_{3}=\frac{\bar{p}_{11}^3-3\tau_{1}\bar{p}_{22}\bar{p}_{11}+2\tau_{2}\bar{p}_{33}}{1-3\tau_{1}+2\tau_{2}},\quad  \notag \\ c_{4}=\frac{\bar{p}_{11}^4-6\tau_{1}\bar{p}_{22}\bar{p}^{2}_{11}+3\tau_{1}^2\bar{p}^{2}_{22}+8\tau_{2}\bar{p}_{33}\bar{p}_{11}-6\tau_{3}\bar{p}_{44}}{1-6\tau_{1}+3\tau^{2}_{1}+8\tau_{2}-6\tau_{3}} \label{eq:stdeqn}
\end{gather}
where particles are taken from $|\eta|<2.5$ and only unique particle combinations in the event are considered. 

In the subevent method (2sub), particle combinations are chosen from two rapidity separated subevents $a$ ($-2.5<\eta<-0.75$) and $c$ ($2.5>\eta>0.75$). The gap between the subevents reduce short-range correlations. The correlators in this case are given by
\begin{gather}
  c_{2,\mathrm{2sub}}=(\bar{p}_{11})_{a}(\bar{p}_{11})_{c},\; c_{3,\mathrm{2sub1}}=\frac{(\bar{p}^{2}_{11}-\tau_{1}\bar{p}_{22})_{a}(\bar{p}_{11})_{c}}{1-\tau_{1a}},\;
  c_{3,\mathrm{2sub2}}=\frac{(\bar{p}^{2}_{11}-\tau_{1}\bar{p}_{22})_{c}(\bar{p}_{11})_{a}}{1-\tau_{1c}},\; c_{4,\mathrm{2sub}}=\frac{(\bar{p}^{2}_{11}-\tau_{1}\bar{p}_{22})_{a}(\bar{p}^{2}_{11}-\tau_{1}\bar{p}_{22})_c}{(1-\tau_{1a})(1-\tau_{1c})} \label{eq:2subeqn}
\end{gather}
where the lettered subscripts in Eq.~\ref{eq:2subeqn} denote the subevents from which particles are taken. Note that there are two different ways of calculating two-subevent $c_3$, and the final results is calculated as average, i.e. $2c_{3,\mathrm{2sub}}=c_{3,\mathrm{2sub1}}+c_{3,\mathrm{2sub2}}$. The validity of these formulae is confirmed by comparison with results obtained from direct loop calculations. 

The cumulants are then calculated by averaging $c_{n}$ over a given ensemble of events. In practise, they are calculated over an unit $\Nch$ bin and are then combined over wider bins. In this paper, we use the following dimensionless definition for cumulants (also called scaled-cumulants),
\begin{gather}
  k_{2}=\frac{\lr{c_{2}}}{\llrr{\pT}^{2}},\quad k_{3}=\frac{\lr{c_{3}}}{\llrr{\pT}^{3}},\quad k_{4}=\frac{\lr{c_{4}}-3\lr{c_{2}}^{2}}{\llrr{\pT}^{4}},\quad
  k_{2,\mathrm{2sub}}=\frac{\lr{c_{2,\mathrm{2sub}}}}{\llrr{\pT}_a\llrr{\pT}_c},\notag \\ k_{3,\mathrm{2sub1}}=\frac{\lr{c_{3,\mathrm{2sub1}}}}{\llrr{\pT}_a^2\llrr{\pT}_c},\quad k_{3,\mathrm{2sub2}}=\frac{\lr{c_{3,\mathrm{2sub2}}}}{\llrr{\pT}_a\llrr{\pT}_c^2},\quad
  k_{4,\mathrm{2sub}} = \frac{\lr{c_{4,\mathrm{2sub}}}-2\lr{c_{2,\mathrm{2sub}}}^{2}-\lr{c_{2}}_{a}\lr{c_{2}}_{c}}{\llrr{\pT}_a^2\llrr{\pT}_c^2}\;. \label{eq:kn}
\end{gather}
Note that the $\lr{c_{2}}_{a}$ and $\lr{c_{2}}_{c}$ denote the $\lr{c_{2}}$ obtained from the standard method in subevent $a$ and $c$, respectively. The final skewness is taken as $2k_{3,\mathrm{2sub}}=k_{3,\mathrm{2sub1}}+k_{3,\mathrm{2sub2}}$.

For HIJING simulation, we generate $pp$, $p$+Pb, and Pb+Pb collisions at $\sqn=5.02$ TeV and Xe+Xe collisions at $\sqn=5.44$ TeV. The particles are chosen from $|\eta|<2.5$ and two $\pT$ ranges: $0.2<\pT<2$ GeV and $0.2<\pT<5$ GeV. The lower $\pT$ range is less sensitive to contribution from short-range correlations. To facilitate direct comparison with experimental measurements, the calculations are carried out as a function of $\Nch$ consisting only of charged hadrons within $|\eta|<2.5$ and $0.5<\pT<5$ GeV. The $\Nch$ range thus obtained is consistent with the choice used in the ATLAS  experiment~\cite{ATLAS:2019peb}.
\section{Results}\label{sec:result}
\begin{figure*}[!h]
\centering
\includegraphics[width=0.8\linewidth]{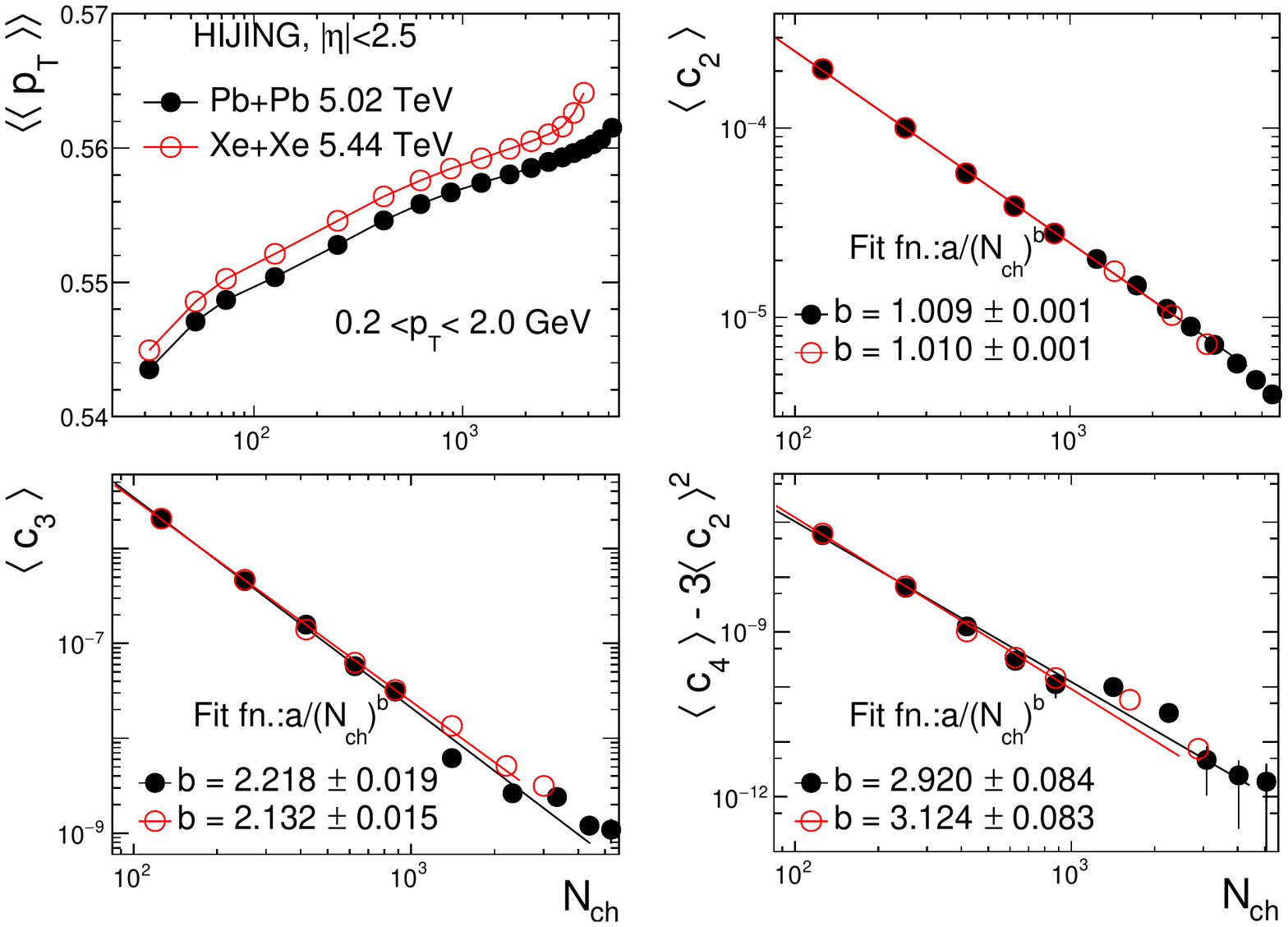}
\caption{ The $\pT$ cumulants without normalizing by $\llrr{\pT}$ for particles in $0.2\space<\pT<\space2$ GeV as a function of $\Nch$ in Pb+Pb and Xe+Xe collisions. The solid lines show the fit of the data to $a/(\Nch)^{b}$. }
\label{fig:powerlawfit}
\end{figure*}

Figure~\ref{fig:powerlawfit} shows the $\pT$ cumulants without the normalization by $\llrr{\pT}$ in Pb+Pb and Xe+Xe collisions as a function of $\Nch$. A double-log scale is used to show clearly the power-law dependence as a function of $\Nch$. The extracted power law index from a simple fit of the form $a/(\Nch)^{b}$ gives the value $b$ that is very close to $(n-1)$ expected for the $n^{\mathrm{th}}-\mathrm{order}$ cumulant. This confirms that the $\pT$ fluctuations in HIJING originate from superposition of independent sources.
\begin{figure*}[!h]
\centering
\includegraphics[width=0.9\linewidth]{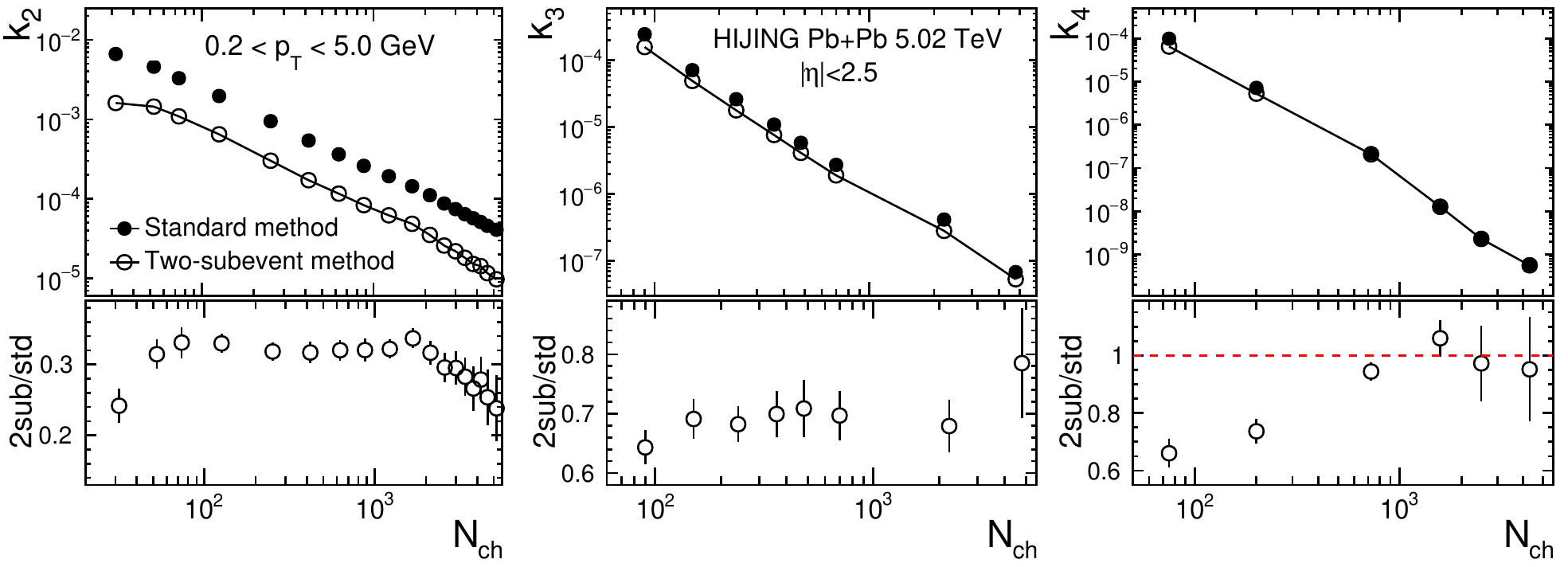}
\caption{The variance (left), skewness (middle) and kurtosis (right) in Pb+Pb collisions 
using the standard (solid points) and two-subevent method (open points) for charged particles in $0.2<\pT<5.0$ GeV as a function of $\Nch$.}
\label{fig:2secorrdep}
\end{figure*}

Figure~\ref{fig:2secorrdep} compares $k_{n}$ between the standard and subevent calculations to estimate the influence of short-range correlations. The $k_{2,\mathrm{2sub}}$ is suppressed by a factor of 3 in comparison to $k_{2}$ from the standard method, suggesting the scaled variance in HIJING model is dominated by the short-range correlations. The influence for $k_3$ and $k_4$ are significantly smaller. This is consistent with the expectation that the short-range correlations have smaller impact for higher-order cumulants.

\begin{figure*}[!h]
\centering
\includegraphics[width=0.9\linewidth]{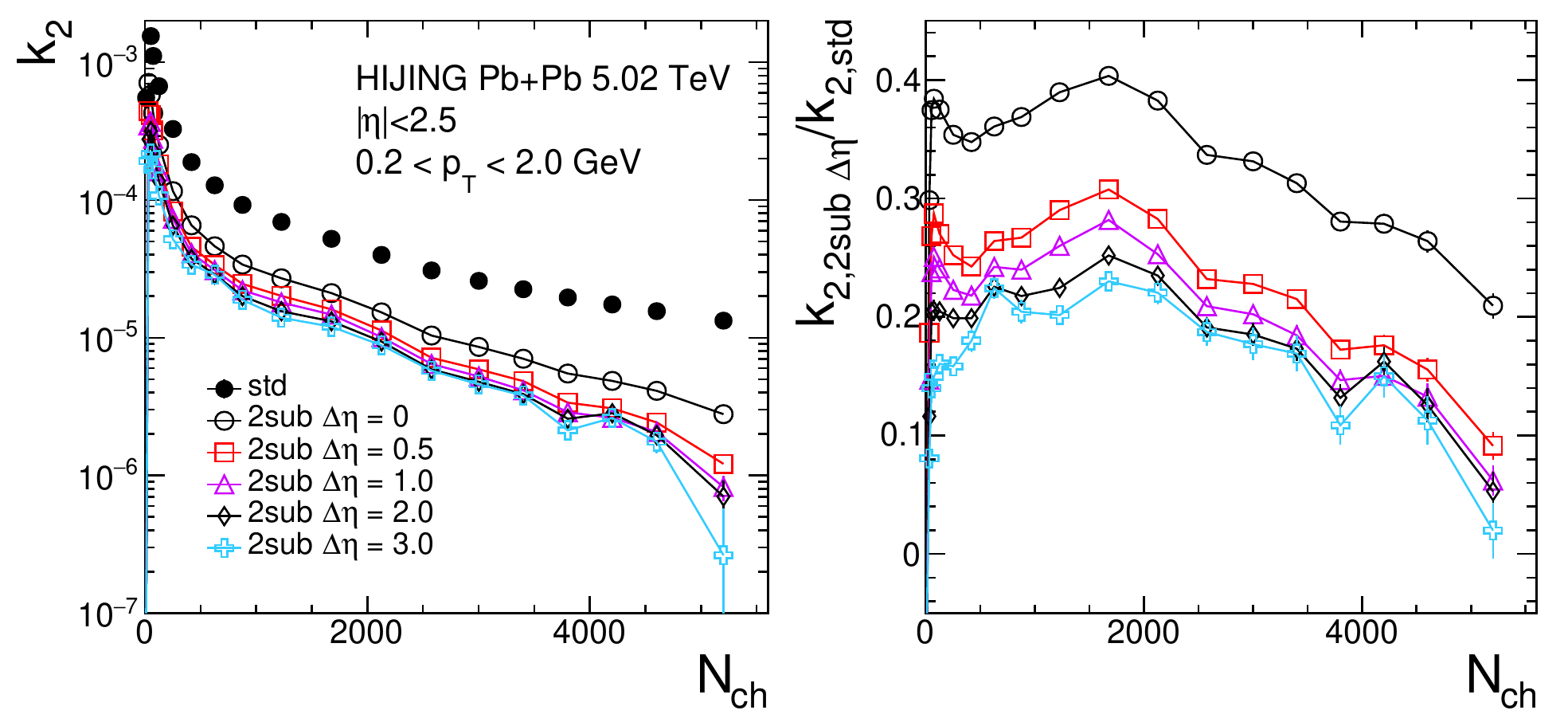}
\caption{(Left) The variance obtained for the standard method and two-subevent method with various rapidity separations in Pb+Pb collisions for $0.2<\pT<2$ GeV as a function of $\Nch$. (Right) Ratio of results from two-subevent method to those obtained from the standard method.}
\label{fig:decor}
\end{figure*}
\begin{figure*}[!h]
\centering
\includegraphics[width=0.8\linewidth ]{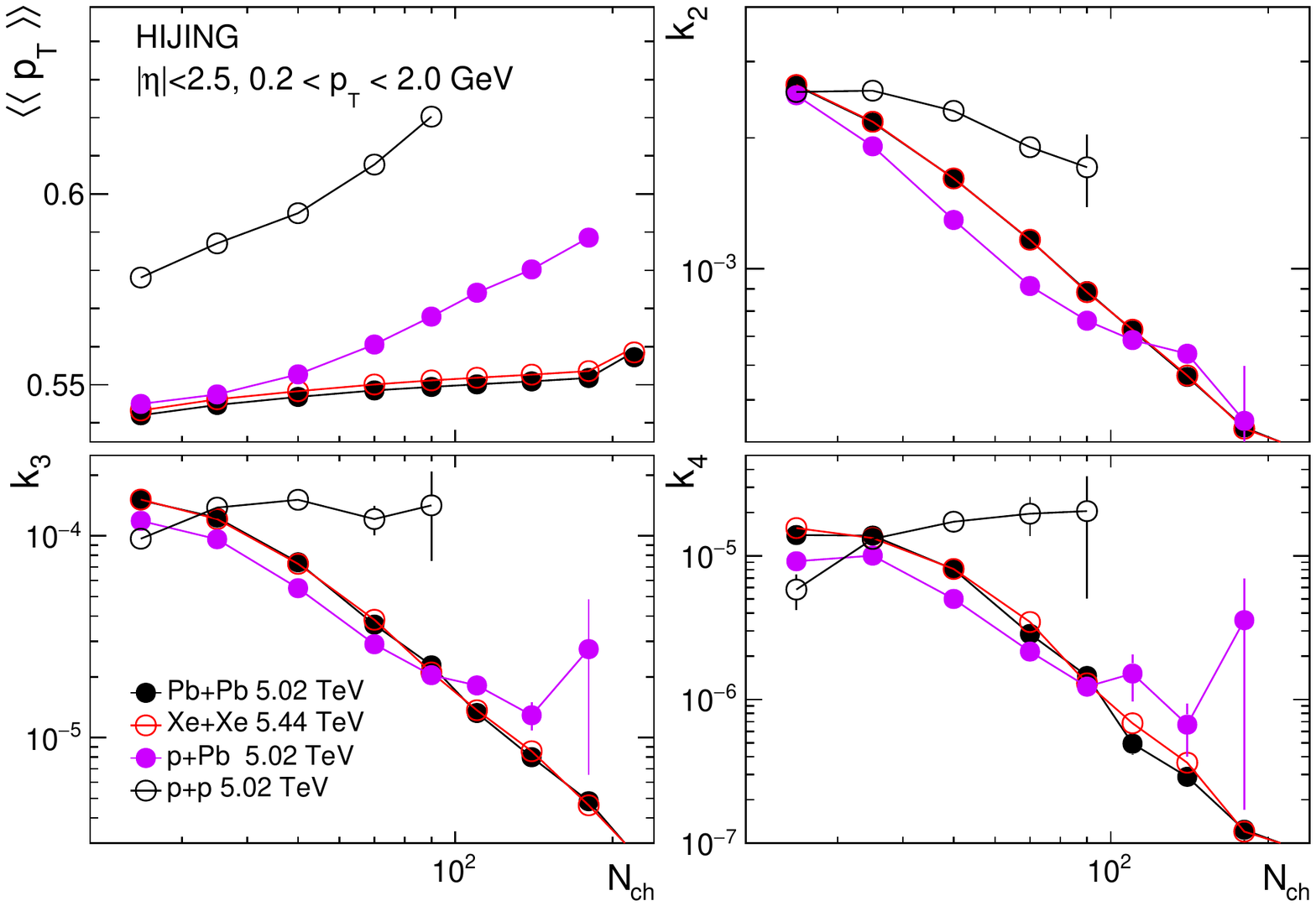}
\caption{The $\pT$ cumulants compared between different collision systems as a function of $\Nch$ }
\label{fig:sysdepcumu}
\end{figure*}

Figure~\ref{fig:decor} shows  how the scaled variance $k_{2}$ depends on the rapidity separation between the two subevents. Most of the decrease is obtained between the subevent method and standard method. Further decrease is observed when one increase the separation between the two subevents. Beyond the edge gap of 0.5 unit, however, the values of $k_{2}$ do not change further. This residual signal probably reflects the genuine long-range correlations associated with strings in the HIJING model.

Figure~\ref{fig:sysdepcumu} compares the $\pT$ cumulants between $pp$, $p+$Pb, Xe+Xe and Pb+Pb collisions as a function of $\Nch$. In HIJING, we expect fluctuations to decrease with increasing system size in accordance with increase of the number of sources. The magnitude of $\llrr{\pT}$ shows a clear system size ordering, whereas for higher order cumulants only $pp$ has a higher magnitude than those for $p+$Pb, Xe+Xe and Pb+Pb collisions. The latter is expected since the number of sources in $pp$ is small, and the events with large $\Nch$ is dominated by fluctuations in the particle production in each source, i.e. from minijets, instead of fluctuations in the number of sources. 

\section{Summary}\label{sec:summary}
In summary, we provide a framework for calculating the higher-order dynamical $\pT$ cumulants up to $4^{\mathrm{th}}$ order using the standard and subevent methods. In the HIJING model, the higher-order cumulants are found to follow a simple power-law scaling as a function of charged particle multiplicity $\Nch$, as expected from a simple superposition of independent sources. The subevent method is found to effectively suppress short-range correlations that dominates the variance of the $\pT$ fluctuations; such short-range correlations have smaller influence on skewness and kurtosis of the $\pT$ fluctuations. The $\pT$ cumulants are found to follow a common scaling as a function of $\Nch$, except in the $pp$ collisions where they are influenced by fluctuations of particle production within each source. Our study provides a useful baseline for $\pT$ fluctuations that is based on simple superposition of independent sources and short-range correlations but without final state effects.

{\bf Acknowledgements:} This work is supported by DOE DE-FG02-87ER40331.
\bibliography{hijing}{}
\bibliographystyle{apsrev4-1}
\end{document}